\documentclass[journal]{IEEEtran}
\usepackage{graphicx}
\usepackage{balance}
\usepackage{comment}
\usepackage{cite}
\usepackage{amssymb}
\usepackage[tight,footnotesize]{subfigure}
\usepackage[active]{srcltx}
\usepackage{amsmath}

\newcommand{\norm}[1]{\left\lVert#1\right\rVert}
\graphicspath{{./figs/}}
\usepackage{eurosym}
\usepackage{algorithm}
\usepackage{algorithmic}
\usepackage{multicol}
\usepackage{dsfont}
\usepackage{amsfonts}
\newcommand{\normvec}[1]{\left\lVert#1\right\rVert}
\usepackage{cases}
\usepackage{bm}
\usepackage{xcolor}

\begin{document}

\title{A Cooperative Multiagent Probabilistic Framework for Search and Track Missions}
\author{Savvas~Papaioannou,~Panayiotis~Kolios,~Theocharis~Theocharides,\\~Christos~G.~Panayiotou~ and ~Marios~M.~Polycarpou 
\thanks{The authors are with the KIOS Research and Innovation Center of Excellence (KIOS CoE) and the Department of Electrical and Computer Engineering, University of Cyprus, Nicosia, 1678, Cyprus. {\tt\small \{papaioannou.savvas, pkolios, ttheocharides, christosp, mpolycar\}@ucy.ac.cy}}}

\markboth{IEEE Transactions on Control of Network Systems, VOL. 8, NO. 2, JUNE 2021}%
{Papaioannou \MakeLowercase{\textit{et al.}}}

\maketitle

\begin{abstract}
In this work a robust and scalable cooperative multi-agent searching and tracking framework is proposed. Specifically, we study the problem of cooperative searching and tracking of multiple moving targets by a group of autonomous mobile agents with limited sensing capabilities. We assume that the actual number of targets present is not known a priori and that target births/deaths can occur anywhere inside the surveillance region thus efficient search strategies are required to detect and track as many targets as possible. To address the aforementioned challenges we recursively compute and propagate in time the \textit{searching-and-tracking} (SAT) density. Using the SAT-density, we then develop decentralized cooperative look-ahead strategies for efficient searching and tracking of an unknown number of targets inside a bounded surveillance area. 
\end{abstract}


\section{Introduction} \label{sec:Introduction}
One of the biggest challenges today’s society faces is its resilience to severe disasters. First responders currently rely on a number of conventional methods to gather information that are time consuming, while the descriptive character of the collected information often lacks accuracy, eloquence and the necessary level of detail. In this work, we envision that a team of autonomous mobile agents (e.g., drones) could become an important technological tool to aid the work of the rescuers. Under this setting, the mission of one or more drone agents is to assist first responders by conducting the following important tasks: a) search the area for situational assessment, and b) detect and track victims as accurately as possible. More specifically, in a cooperative searching and tracking mission, multiple agents are tasked to cooperatively search a certain area of interest in order to discover survivors while at the same time keeping track of those survivors already detected. 
This work builds upon the theory of random finite sets (RFS) and proposes a cooperative multi-agent framework for searching and tracking missions that takes into account the unknown and time varying number of survivors, the noisy sensor measurements and the limited sensing range of the agents. In addition, efficient cooperative searching-and-tracking strategies are devised which allow the agents to generate joint search-plans and detect and resolve tracking overlaps. 

The rest of the paper is organized as follows. Section \ref{sec:Related_Work} reviews the existing literature on the problem of searching and tracking by single and multiple agents. Section \ref{sec:Background} provides a brief overview on multi-target tracking (MTT) and on the theory of random finite sets (RFS). Section \ref{sec:system_model} presents the modeling assumptions of the proposed framework and Section \ref{sec:proposed_approach} presents the details of the proposed approach. Finally, Section \ref{sec:Evaluation} conducts an extensive performance analysis and Section \ref{sec:Conclusion} concludes the paper.

\section{Related Work}\label{sec:Related_Work}

Of particular interest in this work is the problem of cooperative searching and tracking for survivors during search and rescue missions. Previous works in \cite{Bourgault2003} and \cite{Liu2017} investigate the searching and tracking problem but only for the single-agent single-target case. The work in \cite{Furukawa2006} proposes a recursive Bayesian multi-agent searching and tracking solution, however the agents are required to be in communication range at all times. The work in \cite{Frew2008} proposes a task assignment algorithm that integrates area search and target tracking, however requires that the number of agents is larger than the number of targets and that a single agent can only track one target at a time. The problem of multi-agent searching and tracking is also investigated in \cite{Pitre2012} but lacks online path generation. The work in \cite{Peterson2017} proposes a cooperative search and track framework however, requires clutter free environment and perfect target detection. Finally, relevant works also include \cite{Dames2017,Dames2019,Papaioannou2019_1,Papaioannou2019_2} which implement efficient multi-agent RFS-based simultaneous coverage and tracking algorithms for tracking multiple targets. Complementary to the related work, in this paper we propose a decentralized architecture where multiple agents cooperatively search a region of interest in order to detect and track multiple targets.  A preliminary work has been published in \cite{PapaioannouCDC19}. The current work is a more complete study with stronger results.

In this work, we assume that a specific obstacle-free 2D region of interest needs to be continuously searched for potential targets with the aid of a group of mobile agents. The number of targets is not known a priori and may change over time. As a consequence, target births and deaths can occur at random times and the targets can spawn from anywhere inside the surveillance area. The agents are equipped with sensors that are not perfect i.e., as a result of various sensor imperfections, the agents receive noisy target measurements and clutter (e.g., false-alarm measurements). Moreover, the agents have limited sensing range for detecting targets and limited communication range for exchanging information with other nearby agents.  

We should point out that in this work we assume that the mobile agents have perfect self-localization ability and that their dynamical model and control inputs are deterministic.

The objective of each agent at an arbitrary time-step is to: a) accurately estimate the number of targets and their states  from noisy measurements in the presence of clutter, and b) generate search-plans for efficiently searching the whole surveillance area. To achieve a) and b), each agent propagates in time the \textit{searching-and-tracking density} (SAT-density) which is used to a) keep track the areas in the surveillance area that need to be searched and b) estimate the number and states of all targets inside the agent's sensing range, which in this work is achieved with the PHD-filter. 

Moreover, the agents opportunistically cooperate by exchanging information in order to tackle the above objectives more efficiently e.g., two or more agents cooperate to generate joint search-plans and to resolve tracking overlaps (i.e., a situation where 2 or more agents track the same targets). To summarize, the agents opportunistically exchange their search densities, estimated target states and their mode of operation i.e., \textit{searching} or \textit{tracking}. We should also note that all agents operate in \textit{searching} mode optimizing their local or joint search objective(s) (see subsection \ref{ssec:cooperative_search}) until targets are found in the surveillance area in which case the respective agents switch to \textit{tracking} mode (see subsection \ref{ssec:multi_agent_tracking}).

\section{Background} \label{sec:Background}
\subsection{Multi-Target Tracking}

Multi-target tracking (MTT) algorithms aim to estimate the number and states of multiple targets from noisy sensor measurements in the presence of false alarms or clutter and are commonly divided in two main categories namely data-association based and data-association free. Data-association based MTT methods require to first solve the measurements-to-tracks assignment problem and only then proceed with the multi-target state estimation. On the other hand, data-association free methods (e.g., based on random finite sets (RFSs) \cite{Mahler2003}) can bypass the data association problem and proceed directly with the multi-target state estimation. Depending on the application, this property could be highly desirable since it significantly reduces the computational complexity of the MTT algorithms.

Popular data-association based MTT algorithms include multi-scan and single-scan approaches such as the multi-hypothesis tracking algorithm (MHT) \cite{Reid1979} and the joint-probabilistic data-association filter (JPDAF) \cite{Fortmann1983} respectively, sampling-based techniques such as the Rao-Blackwellized Monte Carlo Data Association filter (RBMCDAF) \cite{Sarkka2004} and finally nonparametric Bayesian methods based on Dirichlet processes such as the work in\cite{Fox2006}. We should point out here that not all of the above methods can estimate the number of targets inside the surveillance area. For instance the JPDAF and RBMCDAF algorithms generally require a fixed and known number of targets.  

Data-association free methods are generally based on random finite sets (RFS) and include the probability hypothesis density (PHD) filter \cite{Mahler2003}, the cardinalised PHD 
(CPHD) filter \cite{Mahler2007} and the Multi-Bernoulli filter \cite{Vo2009}. These MTT algorithms can simultaneously estimate the number of targets and their states without solving the data-association problem.

The PHD filter \cite{Mahler2003} is the first practical approximation of the multi-target Bayes posterior which propagates in time the first-order statistical moment instead of the full multi-target distribution. The computationally more expensive CPHD filter \cite{Mahler2007} jointly propagates the first-statistical moment and the cardinality distribution. Unlike the PHD and CPHD filters the multi-Bernoulli filter \cite{Vo2009} approximates the true multi-target posterior distribution as multi-Bernoulli distribution and thus it propagates in time the parameters of a multi-Bernoulli distribution. 

More recently, the RFS MTT approaches have been extended in order to handle the problem of data-association. More specifically, with the introduction of labeled RFSs \cite{Vo2013lrfs}, the generalized labeled multi-Bernoulli (GLMB) filter \cite{Vo2017lrfs} is able to simultaneously estimate the number of targets and their states from a set of noisy observations in the presence of data association uncertainty, detection uncertainty and clutter. This however comes with an increased computational cost compared to the RFS data-association free methods. A more detailed description of MTT algorithms can be found in \cite{VoBook2015}. 

\subsection{Random Finite sets}

A random finite set (RFS) is a finite-set-valued random variable which exhibits the following two properties a) the number of elements in a RFS is random and b) the order of the elements in a RFS is irrelevant. The RFS $X \in \mathcal{F(X})$ is completely specified by a) its cardinality distribution $\rho(n) = p(|X|=n),~ n \in \mathbb{N}_0$ which defines a probability distribution over the number of elements in $X$ and b) by a family of joint probability distributions $p(x_1,...,x_n|n)$,~ $x_1,...,x_n \in \mathcal{X}$ that characterize the distribution of its elements over the state space $\mathcal{X}$. 
Finally, the \textit{multi-target} or (\textit{multi-object}) probability density function (pdf) $f(X)$ of the RFS $X$ is given by: $f(X) = f(\{x_1,...,x_n\}) = n!  \rho(n) p(x_1,...,x_n|n)$. The following RFSs are relevant in this paper:
\subsubsection{Bernoulli RFS} The Bernoulli RFS $X$ can either be empty with probability $1-r, ~ r \in (0,1)$ or be a singleton set (i.e. its set cardinality is equal to one) with (existence) probability $r$ and with its element distributed over the state space $\mathcal{X}$ according to $p(x)$. The Bernoulli multi-object pdf is given by $f(X) =  1-r, \text{if}\> X = \emptyset$, and $ f(X) = r p(x), \text{if }\> X=\{x\}$.
Thus a Bernoulli RFS can be completely characterized by the parameter set $(r, p(x))$.

\subsubsection{Poisson RFS} The poisson RFS $X$ has a cardinality distribution which is Poisson with parameter $\lambda$ i.e., $\rho(n) = \frac{e^{-\lambda}\lambda^n}{n!}, ~n =0,1,2... $ and elements which are independent and identically distributed (IID) random variables and distributed according to $p(x)$ on $\mathcal{X}$. The multi-object pdf is given by $    f(X) = e^{-\lambda}\prod_{x \in X} \lambda p(x)$.


As we have already mentioned the PHD filter \cite{Mahler2003} is an approximation of the multi-target Bayes filter in which the first-order statistical moment or the probability hypothesis density (PHD) is propagated through time instead of the full multi-target distribution. More specifically, the PHD is the conditional density function $D_k(x|Z_{1:k})$ on single targets (objects) $x \in \mathcal{X}$ which when integrated over any region $R \subseteq \mathcal{X}$ of the state space $\mathcal{X}$ gives the expected number of targets (objects) $\hat{n}_k$ contained in $R$ i.e., $\int_{R} D_k(x|Z_{1:k}) dx = \hat{n}_k(R)$.
 
 The multi-target state $\hat{X}_k=\{\hat{x}^1_k,\cdots,\hat{x}^{\hat{n}_k(R)}_k \}$, can be estimated as the $\hat{n}_k$(R) highest local maxima of the PHD. The PHD filter operates recursively in two steps i.e., a time prediction step in which the PHD $D_{k|k-1}(x|Z_{1:k-1})$ is predicted for the next time-step and a measurement update step $D_{k}(x|Z_{1:k})$ in which the received measurement set $Z_k$ at time $k$ is incorporated into the PHD as shown below:
\begin{align} \label{eq:PHDrecur}
& D_{k|k-1}(x|Z_{1:k-1}) = b_{k}(x) ~+ \\ 
&\>\>\>\>\>\>\>\>\>\>\>\>\>\>\>\>\>\>\>\>\>\>\>\>\>\>\>\> \int_{\mathcal{X}} p_S(x^\prime) \pi_{k|k-1}(x|x^\prime) D_{k-1}(x^\prime|Z_{1:k-1}) d x^\prime \notag \\
& D_{k}(x|Z_{1:k}) = \!\!\Big[1 - p_D(x) \Big]D_{k|k-1}(x|Z_{1:k-1}) ~+  \\
&\>\>\>\>\>\>\>\>\>\>\>\>\>\>\>\>\>\>\>\>\>\>\>\>\>\>\>\> \Bigg[ \underset{z \in Z_k}{\sum} \frac{p_D(x) g_k(z|x)}{\kappa(z)+\tau(z)} \Bigg]  D_{k|k-1}(x|Z_{1:k-1}) \notag
\end{align}

\noindent where $b_{k}(x)$ is the PHD of target births, $p_S(x)$ is the probability that a target with state $x$ will survive in the next time step, $\pi_{k|k-1}(x|x^\prime)$ is the single-target transition density, $p_D(x)$ is the target detection probability, $g_k(z|x)$ is the measurement likelihood function, $\kappa(z) = \lambda_c f_c(z)$ is the PHD of false alarms with rate $\lambda_c$ and spatial density $f_c(.)$ and finally $\tau(z) = \int p_D(x^\prime) g_k(z|x^\prime) D_{k|k-1}(x^\prime|Z_{1:k-1}) d x^\prime $.

\color{black}

\section{System Model} \label{sec:system_model}

\subsection{Single Target Dynamics and Measurement Model}\label{ssec:single_target_dynamics}
Let the state of a single target is given by $\bm{x} = (x,\ell) \in \mathcal{X} \times \{0,1\}$, 
where $x \in \mathcal{X}$ is the kinematic state of the target, $\mathcal{X} \subseteq R^{n_x}$ denotes the kinematic state space of the target, $n_x$ is the dimension of the state vector $x$ and $\ell \in \{0,1\}$ is the target label taken from the discrete label space $\{0,1\}$. We denote a true target with the label $\ell=1$ and a virtual target with the label $\ell=0$. True targets represent physical targets inside the surveillance region whose kinematic state $x$ needs to be estimated from a sequence of noisy measurements whereas virtual targets represent static and deterministic locations in the environment (these locations will be used as indicators to show whether specific regions in the area have been searched by the agents). Throughout this paper, the kinematic state spaces of true and virtual targets will be denoted as $\mathcal{X}^1$ and $\mathcal{X}^0$, respectively. The single target kinematic state vector $x_k, k \in \mathbb{N}$  evolves in time according to the following equation:
\begin{subnumcases}{x_k=}
   \zeta(x_{k-1}) + w_k & \text{if} $~x_{k-1} \in \mathcal{X}^1$ \label{eq:single_dynamics_a} \\
   x_{k-1} & \text{if} $~x_{k-1} \in \mathcal{X}^0$ \label{eq:single_dynamics_b}
\end{subnumcases}

\noindent where the function $\zeta : \mathbb{R}^{n_x} \rightarrow \mathbb{R}^{n_x}$ models the dynamical behavior of the target. Eqn. (\ref{eq:single_dynamics_a}) describes the evolution of the state vector as a first order Markov process with transitional density $\pi_{k|k-1}(x_k|x_{k-1}) = p_w(x_k - \zeta(x_{k-1}) )$. The process noise $w_{k} \in \mathbb{R}^{n_x}$ is independent and identically distributed (IID) according to the probability density function $p_w(.)$. In this paper we assume that the kinematic state vector $x_k \in \mathcal{X} \subseteq \mathbb{R}^4$ is composed of position and velocity components in Cartesian coordinates i.e., $x_k = [\text{x},\dot{\text{x}},\text{y},\dot{\text{y}}]^\top$. Since a virtual target is static, its kinematic state vector is of the form $x_k = [\text{x},0,\text{y},0]^\top$.

 When an agent detects a true target i.e., $x_k \in \mathcal{X}^1$ at time $k$, it receives a measurement vector $z_k \in \mathcal{Z}$ (range and bearing observations) which is related to the target kinematic state as follows: $z_k = h(x_k,s_k) + v_k$ where the function $h(x_k,s_k)$ projects the state vector to the measurement space, $s_k$ is the state of the agent at time $k$ (described in the next sub-section) and the random process $v_{k} \in \mathbb{R}^{n_z}$ is IID, independent of $w_k$ and distributed according to $p_v(.)$. Thus, the probability density of measurement $z_k$ for a target with kinematic state $x_k$ when the agent is at state $s_k$ is given by the measurement likelihood function $g_k(z_k|x_k,s_k) = p_v(z_k - h_k(x_k,s_k))$.

On the other hand, virtual targets ($x_k\in\mathcal{X}^0$) represent fixed and known locations in the environment i.e., their states are deterministic and predetermined. Additionally, in this work we assume that the agent dynamics are deterministic and that the agents can perform perfect self-localization (as discussed in subsection \ref{ssec:AgentDynamics}). As a result the states of the virtual targets need not to be estimated. As we discuss in Sec. \ref{sec:proposed_approach} virtual targets are used to keep track whether certain regions inside the surveillance area have been searched or not.  

\subsection{Agent Dynamics} \label{ssec:AgentDynamics}
Let $S = \{1,2,...,|S|\}$ be the set of all mobile agents that we have in our disposal operating in a discrete-time setting. At time $k$, the 2D surveillance region $\mathcal{A} \subseteq \mathbb{R}^2$ is monitored by $|S|$ mobile agents with states $s^1_k,s^2_k,...,s^{|S|}_k$, each taking values in $\mathcal{A}$. Each agent $j$ is subject to the following deterministic dynamics:
\begin{equation} \label{eq:controlVectors}
s^j_{k} = s^j_{k-1} + \begin{bmatrix}
						l_1\Delta_R \text{cos}(l_2 \Delta_\theta)\\
						l_1\Delta_R \text{sin}(l_2 \Delta_\theta)
					\end{bmatrix},  
					\begin{array}{l} 
						l_2 = 0,...,N_\theta-1\\ 
						l_1 = 0,...,N_R
				    \end{array} 
\end{equation}
where  $s^j_{k-1} = [s^j_x,s^j_y]^\top_{k-1}$ denotes the position (i.e., $(x, y)$ coordinates) of the $j_{\text{th}}$ agent at time $k-1$, $\Delta_R$ is the radial step size, $\Delta_\theta=2\pi/N_\theta$ and the parameters $(N_\theta,N_R)$ specify the number of possible control actions. We denote the set of all admissible control actions of agent $j$ at time $k$ as $\mathbb{U}^j_{k}=\{s^{j,1}_{k},s^{j,2}_{k},...,s^{j,|\mathbb{U}_{k}|}_{k}  \}$ as computed by Eqn. (\ref{eq:controlVectors}). We should point out that the agent dynamical model is noise-free and that the agents have perfect self-localization ability (i.e., through a very accurate GPS system). 

\subsection{Single Agent Sensing Model} \label{ssec:sensing_model}
The ability of an agent to sense true targets inside the surveillance area is modeled by the function $p_D(x_k,s_k)$ that measures the probability that a target with kinematic state $x_k \in \mathcal{X}^1$ at time $k$ is detected by an agent with state $s_k$. More specifically, the sensing capability of the agent is given by:
\begin{equation}\label{eq:sensing_model_true}
 p_D(x_k \in \mathcal{X}^1 ,s_k) = 
  \begin{cases} 
   p_D^\text{max} & \text{if } x_k \in \mathcal{S}_a(s_k) \\
   0 & \text{if } x_k \notin \mathcal{S}_a(s_k)
  \end{cases}
\end{equation}
where $\mathcal{S}_a(s_k)$ denotes the agent's sensing area which in this work includes all $(x, y)$ points that satisfy the equation $\max \{|x-s_x|,|y-s_y|\}\leq\frac{a}{2}$; i.e., a square region with total area $a^2$ units, centered at $s_k = [s_x, s_y]^\top$ and $p_D^\text{max}$ denotes the probability that the agent's sensor detects the true targets located inside its sensing range. Although, in this work we use a square region to model the agent's sensing area, the proposed approach is not limited to square sensing areas, for instance rectangular and circular sensing areas can also be used.

On the other hand, the agent with state $s_k$ observes virtual targets $x_k \in \mathcal{X}^0$ and determines if they reside within its sensing area using the following indicator function: $\mathds{1}_{\mathcal{S}_a}(x_k \in \mathcal{X}^0 ,s_k) = 1$ if $x_k \in \mathcal{S}_a(s_k)$ and 0 otherwise.
Finally, any two agents with states $s^i_k$ and $s^j_k$ are able to communicate with each other when $\normvec{s^i_k - s^j_k}_2 \le C_R$ where $C_R$ is the communication range with $C_R \geq \sqrt{2}\frac{\alpha}{2}$ i.e., the communication range is greater than or equal to the sensing range to allow for multiple agents to exchange information to avoid tracking overlaps (see subsection \ref{ssec:multi_agent_tracking}).

\subsection{Multi-object dynamics and measurement models}
Multiple independent true targets can exist and evolve inside the surveillance region. True targets (i.e., with label $\ell=1$) can spawn from anywhere inside the surveillance region and target births and deaths occur at random times. This means that at each time $k$, there exist $n^{\ell=1}_k$ true targets with kinematic states $x^1_k, x^2_k,...,x^{n^{\ell=1}_k}_k$, each taking values in the state space $\mathcal{X}^1$, where both the number of true targets $n^{\ell=1}_k$ and their individual states $x_k^i, \forall i \in n^{\ell=1}_k$ are random and time-varying. The RFS of the multi-target state of the true targets $X^{\ell=1}_k \in \mathcal{F(X^\text{1})}$ evolves in time according to: $ X^{\ell=1}_k = [\underset{x_{k-1} \in X^{\ell=1}_{k-1}}{\cup} \Psi(x_{k-1})]  \cup B_k$
where $X^{\ell=1}_{k-1}$ is the multi-target state of the true targets of previous time-step, $\Psi(x_{k-1})$ is a Bernoulli RFS which models the evolution of the set from the previous state, with parameters $(p_{S}(x_{k-1}),\pi_{k|k-1}(x_k|x_{k-1}))$. Thus a target with kinematic state $x_{k-1}$ continues to exists at time $k$ with surviving probability $p_{S}(x_{k-1})$ and moves to a new state $x_k$ with transition probability $\pi_{k|k-1}(x_k|x_{k-1})$. Otherwise, the target dies with probability $1-p_{S}(x_{k-1})$. The term $B_k$ is a Poisson RFS of spontaneous target births.

At time $k$, an agent receives a finite set of measurements from the detected true targets and from clutter denoted as $Z_k$. The multi-target measurement set is formed according to: $Z_k = [ \underset{x_{k} \in X^1_{k}}{\cup} \Theta(x_{k}) ] \cup \text{K}_k$
where $\Theta(x_{k})$ is a Bernoulli RFS with parameters $(p_{D}(x_k,s_k),g_k(z_k|x_k,s_k))$. Thus a true target with kinematic state $x_k$ at time $k$ is detected by the agent with state $s_k$ with probability $p_{D}(x_k,s_k)$. This agent then receives a measurement $z_k$ with likelihood $g_k(z_k|x_k,s_k)$. The target is not detected with probability $1-p_{D}(x_k,s_k)$ and no measurement is being received by the agent. Additionally, an agent can receive false alarms measurements i.e., the term $\text{K}_k$ is a Poisson RFS which models the set of false alarms or clutter received by an agent at time $k$ with PHD $\kappa_k(z_k) = \lambda_c f_{c}(z_k)$, where in this paper $f_c(.)$ denotes the uniform distribution over $\mathcal{Z}$ and $\lambda_c$ is the average number of clutter generated measurements per time-step. 

The virtual targets i.e., $(\ell=0)$ on the other hand do not exhibit any birth and death events. Their number $n^{\ell=0}$ is fixed and their states are known. Thus the multi-target state of virtual targets is given by 
$X^{\ell=0}=\{ x^1,\cdots,x^{n^{\ell=0}}\}, x \in \mathcal{X}^0$. In essence the multi-target state of virtual targets represents discrete locations inside the surveillance environment which in turn correspond to regions for which a search value is computed. The search value shows whether or not these regions have been searched by the agent. Additionally, we should point out here that the state-spaces of true and virtual targets are distinct and thus all true targets automatically received the label $l=1$ whereas virtual targets receive $l=0$.

\section{Proposed Approach} \label{sec:proposed_approach}
In this section we first describe how the proposed approach recursively propagates in time the SAT-density and then we discuss how using the SAT-density the agents cooperate to produce joint search-plans and resolve tracking overlaps. 

\subsection{Searching-and-Tracking Density}
During a search and track mission, a single agent is required to be able to perform the following tasks: a) simultaneously estimate the time-varying number of targets and their states from a sequence of noisy measurements, and b) search the surveillance region for targets as efficiently as possible.

The first task can be accomplished by recursively computing and propagating in time, the PHD of the full multi-target posterior distribution of the true targets using the PHD filter \cite{Mahler2003}. In order to accomplish the second task, the agent needs to: a) keep track of the visited (i.e., searched) and unvisited regions of the surveillance area, b) estimate when and how often certain search regions need to be revisited and c) generate efficient search plans for searching the area. 

In order to perform the aforementioned tasks, the agent stores a discrete representation of the environment in its memory in the form of a graph $G=\{\mathcal{V},\mathcal{E}\}$ termed as \textit{search map}, where each node $v \in \mathcal{V}$ corresponds to a region $r_v \subset \mathcal{A}$ in the surveillance area where $\bigcup_v r_v = \mathcal{A}$. The nodes $v$ of this graph correspond to virtual targets i.e., the location of each virtual target $x \in X^0$ is represented by a node $v \in \mathcal{V}$ with $|\mathcal{V}|=|X^0|$. The agent recursively computes the \textit{search value} $p^\text{search}(r_v) \in [0,1], v\in \mathcal{V}$ of each region and uses this information to decide how often to visit a particular region and how to generate search-plans for efficiently searching the surveillance area. 

With this in mind, we can now discuss how the agent recursively computes the \textit{searching-and-tracking density} or SAT-density. In essence the SAT-density includes two components: a) the search density of the surveillance area and b) the target density inside the agent's sensing range. The search density is used to indicate which areas in the environment have been searched and how often certain regions need to be searched, whereas the target density is used to estimate the number and states of the targets inside the agents sensing range. 

The SAT-density denoted as $\tilde{D}_{k}(\bm{x}|Z_{1:k}) ~\text{for}~ \bm{x} \in  \mathcal{X} \times \{0,1\}$ is a compound density which includes the target density on true targets and the search density on virtual targets. The SAT-density is recursively computed using a prediction and update step i.e., $\tilde{D}_{k-1}(\bm{x}|Z_{1:k-1}) \rightarrow \tilde{D}_{k|k-1}(\bm{x}|Z_{1:k-1}) \rightarrow \tilde{D}_{k}(\bm{x}|Z_{1:k})$. In what follows we denote the target density on true targets as $\tilde{D}_{k}(x \in \mathcal{X}^1)$ and the search density on virtual targets as $\tilde{D}_{k}(x \in \mathcal{X}^0)$ where the dependance on the measurements $Z_{1:k}$ is dropped for notational convenience. The prediction step is given by:
\begin{align}
&\tilde{D}_{k|k-1}(x \in \mathcal{X}^1) = b_{k}(x) ~+ \label{eq:p1} \\ 
&\>\>\>\>\>\>\>\>\>\>\>\>\>\>\>\>\>\>\>\>\>\>\>\>\>\>\> \int_{\mathcal{X}^1} p_S(x^\prime) \pi_{k|k-1}(x|x^\prime) \tilde{D}_{k-1}(x^\prime) d x^\prime  \notag \\
&\tilde{D}_{k|k-1}(x \in \mathcal{X}^0) = \tilde{D}_{k-1}(x\in \mathcal{X}^0) ~ \times \label{eq:p2} \\
&\>\>\>\>\>\>\>\>\>\>\>\>\>\>\>\>\>\>\>\>\>\>\>\>\>\>\> \Big[ \big(1 -  \mathds{1}_{\mathcal{S}_a}(x ,s_{k-1})\big)  J_k(x) +  \mathds{1}_{\mathcal{S}_a}(x ,s_{k-1}) \Big]  \notag 
\end{align}

\noindent where Eqn. (\ref{eq:p1}) is due to the PHD filter i.e., the prediction of the probability hypothesis density of the true targets before the collection of target measurements. On the other hand, Eqn. (\ref{eq:p2}) computes the predicted search density on the state space of virtual targets where the term $\tilde{D}_{k-1}(x\in \mathcal{X}^0)$ is the search density of the previous time step, the indicator function $\mathds{1}_{\mathcal{S}_a}(x ,s_{k-1})$ was defined in subsection \ref{ssec:sensing_model} and finaly $J_k(x) \in [0,1]$ is a function that determines the decay value of the virtual target with state $x$. In essence, the states of all virtual targets outside the agent's sensing range are adjusted accordingly to reflect the fact that they are not being observed. This property is used to generate search plans which will guide the agent to visit areas that have not been recently visited. The updated SAT-density is given by:
\begin{align} 
& \tilde{D}_{k}(x\in \mathcal{X}^1) = \Big[1 - p_D(x, s_k) \Big] \tilde{D}_{k|k-1}(x\in \mathcal{X}^1) ~+ \label{eq:u1} \\
& \>\>\>\>\>\>\>\>\>\>\>\>\>\>\>\>\>\>\>\>\>\>\>  \underset{z \in Z_k}{\sum} \frac{p_D(x, s_k) \cdot g_k(z|x,s_k)}{\kappa(z)+\tau(z)} \tilde{D}_{k|k-1}(x\in \mathcal{X}^1) \notag \\
&\tilde{D}_{k}(x\!\!\in\!\! \mathcal{X}^0)\!\! =\!\!\frac{\mathds{1}_{\mathcal{S}_a}(x ,s_{k})}{|\mathcal{A}|}\!\! + \!\!\Big[1\!\!-\!\!\mathds{1}_{\mathcal{S}_a}(x ,s_{k})\Big] \tilde{D}_{k|k-1}(x\!\!\in\!\! \mathcal{X}^0) \label{eq:u2}
\end{align}

\noindent In the above, Eqn. ({\ref{eq:u1}}) is the update step of the PHD filter, in which the received measurement set $Z_k$ is used to compute the posterior probability hypothesis density of the true targets whereas Eqn. ({\ref{eq:u2}}) is used to update the search density of virtual targets. In essence the search density is adjusted inside the agents sensing range to account for the agent's updated position $s_k$.
\color{black}
The search value $p^\text{search}_k(r_v)$ of a particular region $r_v \subset \mathcal{A},\>v\in\mathcal{V}$ can be computed by integrating the search density in $r_v$ as follows:
\begin{equation} \label{eq:search_value}
p^\text{search}_k(r_v)=\frac{\int_{r_v}\tilde{D}_k(x \in \mathcal{X}^0) dx}{|r_v||\mathcal{A}|^{-1}}
\end{equation}
where $|r_v|$ is the total area of region $r_v$. On the other hand, the number of true targets $\hat{n}_k$ (rounded to the nearest integer) inside the area $R \subseteq \mathcal{A}$ can be computed by integrating the target density in $R$ as $\hat{n}_k(R) = \int_{R} \tilde{D}_k(x \in \mathcal{X}^1) dx$ and the multi-target state $X^1_k$ can be estimated by finding the $\hat{n}_k(R)$ highest peaks  of the PHD. This procedure is due to the PHD filter. 


\subsection{Multi-agent Searching} \label{ssec:cooperative_search}
The search objective is to find the optimal control actions that will move the agent along areas that have not been explored for some time and could potentially reveal new targets. To address this challenge, we first discuss how searching takes into account the search map derived from the SAT-density and how low level controls employ the computed paths to steer the agent across the field. 

\textbf{a) Search planning:} Given the search map $G=(\mathcal{V},\mathcal{E})$ where the set of edges in $\mathcal{E}$ connect adjacent nodes, the cost $c_{il}$ on edge $(i,l)$ is defined as the Euclidean distance between the particular adjacent regions in the field. For each node on this graph, a search value $p^{\text{search}}(r_v),\>v\in\mathcal{V}$ is computed using Eqn. (\ref{eq:search_value}). This value varies between 0 and 1, where 0 indicates that the particular node (and hence region in the field) has not been searched and 1 indicates that the region has just been visited. The SAT-density recursion in Eqn. (\ref{eq:p2}) and (\ref{eq:u2}) indicates how the search value decays over time in order to steer agents to revisit particular regions in the field.

Using $p_k^\text{search}$, we then define the set of unvisited nodes $\bar{\mathcal{V}}$ as the set of nodes for which the search value goes below a certain threshold, i.e., $p_k^\text{search}(r_v)\leq\beta, \> v\in\bar{\mathcal{V}}$, thus indicating that those nodes need to be revisited. Given $\bar{\mathcal{V}}$ and the current agent state $s_k$ of an agent, we would like to compute paths that visit nodes in $\bar{\mathcal{V}}$ with the least cost $c_{il}$ in order to search the whole region for targets. 

To compute optimal paths for each agent, we formulate the following integer linear program $(P1)$, with the objective of minimizing the total traversal cost. 
\begin{align}
\mathrm{(P1)}\quad&\min\>\>\sum_{(i,l)\in\mathcal{E}} c_{il} y_{il} \label{eq:objective_P1}\\
\mathrm{s.t.} \quad & \sum_{l: (i,l) \in\mathcal{E}} y_{il} \geq 1, \quad \forall\> i\in\bar{\mathcal{V}} \label{eq:nodes_visited}\\
& \sum_{l: (i,l) \in\mathcal{E}} y_{il} - \sum_{l: (l,i) \in\mathcal{E}} y_{li} = 0, \quad \forall\> i\in\mathcal{V} \label{eq:conservation_of_flow}\\
&\sum_{l:(s_k,l) \in\mathcal{E}} y_{s_kl} \geq 1 \label{eq:connected_to_s}\\
& M\!\!\! \sum_{(i,l) \in \mathcal{E}(\mathcal{Q})} y_{il} \geq \!\!\!\!\!\sum_{(i,l) \in \mathcal{E}'(\mathcal{Q})} y_{il}, \quad \forall\> \mathcal{Q} \subseteq \mathcal{V} \setminus \{s_k\} \label{eq:ASEC}\\
& y_{il}\in\mathbb{Z}_{\geq 0}, \quad \forall\> (i,l) \in \mathcal{E} \label{eq:domains}
\end{align}

\noindent The integer variable $y_{il}$ in $(P1)$ indicates the number of times that the agent traverses the particular edge $(i,l)$. The objective function in Eqn. (\ref{eq:objective_P1}) ensures that the least-cost path is selected. Constraint (\ref{eq:nodes_visited}) ensures that all nodes in the unvisited list $\bar{\mathcal{V}}$ are visited at least once. Note that through this constraint, we allow each agent to visit a node more than once if that is necessary to minimize the total route cost. The conservation of flow constraints at each node $i$ are given in constraint equations (\ref{eq:conservation_of_flow}). To accommodate an open search path, the flow-balance in Eqn. \eqref{eq:conservation_of_flow} can be extended by setting the right-hand side to $\mathds{1}_{\{i=s_k\}} - e_i$. This uses a binary indicator $e_i \in \{0,1\}$ to dynamically designate an optimal free terminal node, where network flow principles intrinsically guarantee $\sum_{i \in \mathcal{V}} e_i = 1$. In order to ensure that the computed path is connected we impose constraints \eqref{eq:connected_to_s} and \eqref{eq:ASEC}. The former constraint ensures that any resulting path starts from state $s_k$ (allowing for multiple passes), and the latter constraint strictly prevents disconnected flow cycles (subtours). In constraint \eqref{eq:ASEC}, for any non-empty subset of nodes $\mathcal{Q} \subseteq \mathcal{V} \setminus \{s_k\}, \mathcal{Q}\neq \emptyset$ that explicitly excludes the starting state, we define $\mathcal{E}(\mathcal{Q})$ to be the set of edges with exactly one end in $\mathcal{Q}$ (the cut-set boundary) and let $\mathcal{E}'(\mathcal{Q})$ to be the set of internal edges with both ends in $\mathcal{Q}$. The right hand side of the constraint is the total flow internal to $\mathcal{Q}$ and the sum on the left hand side is the total flow crossing the boundary of $\mathcal{Q}$. Thus if there is some internal flow in $\mathcal{Q}$ then the right hand side is strictly positive which forces the left hand side to be positive, guaranteeing that the flow in $\mathcal{Q}$ is connected to the main tour originating from $s_k$.

When solved to optimality, $(P1)$ computes the best alternative path at the minimum cost. However due to the exponential number of constraints in Eqn. (\ref{eq:ASEC}), $(P1)$ is computationally hard to solve in practice. Hence, an alternative heuristic approach is followed in the sequel to devise a path considering the constraints as expressed in $(P1)$. Starting at $s_k$, paths are built by adding new edges with the least cost $c_{il}$, beginning from the head node of the last edge added. The process terminates when all nodes have been visited or when no more edges can be added.

\textbf{b) Search Control:} Given the computed path sequence, the objective is then to take a control action $u_k \in \mathbb{U}_k$ that will move each agent across the designated path. To achieve this, a list $\bar{\mathcal{V}}$ of nodes to-be-visited is maintained, and each node is marked as visited whenever the agent moves to a position where the particular node is within its sensing range. The search control objective can be expressed as follows: $u_{k}^{\star} = \underset{u_k}{\arg\min}~ \xi_\text{search}(u_{k},v)$ where $v\in\bar{\mathcal{V}}$ indicates the location of the next unvisited node in the list and $\xi_\text{search}$ returns the Euclidean distance between the position of the agent (when the hypothetical control action $u_k$ is applied) and the next unvisited node $v$. By iteratively visiting the path sequence, the envisioned look-ahead search control is achieved by each agent.

\textbf{c) Search Cooperation:} Whenever two or more agents are in communication range they exchange their search densities and merge their copies using a simple max operation of local and received values. A fused search map is then computed which contains the search-path histories of the cooperating agents. The agents can then compute a joint search-plan as follows: Let $\bar{S}$ be the subset of agents in communication range and assume that each agent knows the number $|\bar{S}|$ and position $s_k^j,\>j\in \bar{S}$ of cooperating agents in its vicinity. Iteratively, each agent computes $|\bar{S}|$ paths incrementally by adding one node at a time in each agent's path from the list of all unvisited nodes in $\bar{\mathcal{V}}$, until there are no more unvisited nodes. A new node is added in an agent's path only if the head node of all possible edges to traverse (starting from the edge with the least cost) is not flagged as visited and the tail node of that edge is the last node added in the particular walk. The steps followed by each agent are detailed in Algorithm \ref{alg:oppsearch}.



\begin{algorithm}
	\caption{Opportunistic multi-agent search cooperation}\label{SOC}
	\begin{algorithmic}
		\REQUIRE $|\bar{S}|$ and $s_k^j,\>j\in \bar{S}$, $G=(\mathcal{V},\mathcal{E})$ and let $\bar{\mathcal{V}}=\mathcal{V}$
		\STATE {$T_j \gets \{s_k^j\}$} \COMMENT{$T_j$ contains the path of agent $j\in \bar{S}$}
		\STATE Set $\bar{\delta}=0$ \COMMENT{Indicating the tree to be appended first}
		\WHILE{($\bar{\mathcal{V}} \neq \emptyset$)}
		\STATE $\bar{\delta}=\bar{\delta}+1, j= \bar{\delta}\>\text{mod}\> |\bar{S}|$
		\STATE {$l \gets \arg\min_{v\in\bar{\mathcal{V}}} c(i,v)$} \COMMENT{where $i$ is the last node added in $T_j$}
		\STATE {$T_j\gets T_j \cup \{l\}$} \COMMENT{Update tree $T_j$} 
		\STATE {$\bar{\mathcal{V}} \gets \bar{\mathcal{V}} \setminus \{l\}$} \COMMENT{Remove node $l$ from unvisited set}
		\ENDWHILE		
	\end{algorithmic}
	\label{alg:oppsearch}
\end{algorithm}


We should point out that Alg. \ref{alg:oppsearch}, is a greedy heuristic of the vehicle routing problem (VRP) \cite{Laporte1992}, and ensures that each computed path is minimum with respect to the cost of the edges it traverses on the graph. When executed by each agent the individual minimum cost paths will be computed but these paths are not necessarily globally optimal. Moreover, Alg. \ref{alg:oppsearch} is executed by each agent after the search density has been exchanged and merged among those agents in communication range. Alternatively, the algorithm can be executed by a single agent and then the path plans can be send out to all other agents after the computation has been completed. This however, will entail an extra communication cost. Finally, please note that the search density decays over time according to Eqn. (\ref{eq:p2}) and also captures whole surveillance area. Thus any unvisited regions in the surveillance area (or regions that have not visited recently) would have low search values pushing agents to compute path plans that include those regions and thus exploring the unvisited areas.

\textbf{d) Communication Overhead:} When the agents are in communication range they exchange information (i.e., their search densities) in order to compute joint search plans as already explained. Although, this information exchange increases the searching efficiency, it also creates a communication overhead. Assuming that the search density is implemented in this work as a discrete set of values  i.e., a set of $N$ real numbers forming a grid over the surveillance area, then whenever two agents are in communication range they exchange $2N$ real numbers in total. This is the communication cost in terms of the amount of information that needs to be transmitted.
There are a number of ways which can be employed in order to reduce this communication overhead. In this work, we use the following event based strategy i.e., when two agents are in communication range they exchange their search densities and they compute a joint search plan. However, these two agents refrain to exchange their search densities with each other while executing their joint search plan. Thus the agent $i$ is allowed to exchange information only when a) is not participating in a joint search plan or b) is participating in a joint search plan and encounters an agent $j$ which is not involved in agent's $i$ joint search plan. A more detailed analysis and study of the communication aspects of this work will be investigated in the future. 

\subsection{Multi-agent Tracking} \label{ssec:multi_agent_tracking}
In this subsection we discuss: a) how the agents select control actions in order to accurately track multiple targets and b) how multiple agents are cooperating to detect and resolve tracking overlaps.

\textbf{a) Tracking Control:} The objective of tracking control is to find the optimal control action $u_k \in \mathbb{U}_k$ that must be taken at time step $k$ by each agent in order to maintain tracking of the detected targets. We should point out that the control actions $u_k$ applied to the agents affect the received measurements $Z_k$ which in turn affect the multi-target state estimate $\hat{X}^1_k$ during the update step. In other words the received measurement set $Z_k$ (if any) depends on which control action $u_k$ has been applied (e.g., a target might not be detected if the wrong control action is applied). This can be seen from the agent's sensing and measurement models as discussed in Sec. \ref{sec:system_model}. Thus, ideally to optimize the control actions, would require the knowledge of the future measurement set $Z_k$. 
Since, the future measurement set $Z_k$ is not available until the control action $u_k$ is applied, we generate the predicted measurement set $\bar{Z}_{k}$ and we use it in place of $Z_k$ in order to optimize the above objective function. More specifically, the predicted measurement set for each control action is generated as follows: $\bar{Z}_{k} =~ \bar{Z}_{k} ~ \cup ~\{{\arg\max}_z ~  g_k(z|x,u_k)\}$ for all $x \in \hat{X}^1_{k|k-1}$ and for all $u_k \in \mathbb{U}_k$
where $\hat{X}^1_{k|k-1}$ is the predicted multi-target state for the true targets, which can be obtained by integrating the predicted target density using Eqn. (\ref{eq:p1}) to compute the predicted number of targets and then extracting their states from the PHD. Let the tracking objective function be denoted as $\xi_\text{track}(u_k,Z_k)$. Using the predicted measurement sets $\bar{Z}_{k}$ in place of the actual measurement set, the control problem becomes: $u_{k}^{\star} = \underset{u_k}{\arg\max} ~  \xi_\text{track}(u_{k},\bar{Z}_{k})$. To optimize the tracking objective, the following steps are performed: For each admissible control action $u_k \in \mathbb{U}_k$ we generate the predicted measurement set $\bar{Z}_{k}$. For each pair $(u_k, \bar{Z}_{k})$ we perform a pseudo-update step using Eqn. (\ref{eq:u1}) to produce the (pseudo) posterior target density which we denote as $\bar{D}_k(x \in \mathcal{X}^1)$. We consider the information gain between the predicted $f_{k|k-1}(X|Z_{1:k-1})$ and the (pseudo) updated $\bar{f}_k({X|Z_{1:k-1},\bar{Z}_{k},u_k})$ multi-target distributions as a measure of decreasing the uncertainty of the estimated multi-target state. The objective is then to maximize the information gain between the two multi-target distributions. To measure the information gain, we use as $\xi_\text{track}(u_k, \bar{Z}_{k})$ the Renyi divergence which is given by:
\begin{align}\label{eq:divergence}
     \frac{1}{\alpha-1} \text{log}\!\! \int [\bar{f}_k(X|\bar{Z}_{k},u_k)]^\alpha [f_{k|k-1}(X|Z_{1:k-1})]^{1-\alpha}\!\delta X
\end{align}
where $0 < \alpha < 1$ determines the emphasis given on the tails of the two distributions \cite{Ristic2011}. Finally, Eqn. (\ref{eq:divergence}) becomes for our problem:
\begin{align} \label{eq:track_control}
& \int_{\mathcal{X}^1} \tilde{D}_{k|k-1}(x) dx + \frac{\alpha}{(1-\alpha)} \int_{\mathcal{X}^1} \bar{D}_{k}(x|\bar{Z}_{k},u_k) dx ~- \notag  \\
& \frac{1}{(1-\alpha)} \int_{\mathcal{X}^1} \bar{D}_{k}(x|\bar{Z}_{k},u_k)^\alpha \tilde{D}_{k|k-1}(x)^{1-\alpha} dx
\end{align}
where  $\tilde{D}_{k|k-1}(x \in \mathcal{X}^1)$ is the predicted target density according to Eqn. (\ref{eq:p1}) and $\bar{D}_k(x \in \mathcal{X}^1|\bar{Z}_{k},u_k)$ is the (pseudo) updated target density according to Eqn. (\ref{eq:u1}) in which we show explicitly its dependance on the hypothetical control action ($u_k$) and predicted measurement set ($\bar{Z}_{k}$). To summarize, the agent selects the control action (before actually receiving the measurement) which makes the divergence between the predicted and updated multi-target densities as large as possible i.e., maximizing the amount of information in the posterior multi-target density with respect to the predicted density.

\textbf{b) Tracking Cooperation:} In this paragraph we discuss how our approach can handle tracking overlaps i.e., a problem in which a target (or a group of targets) is being tracked by more than one agent. The single agent tracking control strategy discussed in the previous paragraph can cause this problem, which in this work is something undesirable since valuable system resources are wasted for performing the same task i.e., tracking the exact same targets. The objective of the tracking overlap detection and resolution is to maximize the utilization of resources. 
In particular, consider the scenario where 3 targets, which are being tracked by 2 different agents, approach each other over time. Eventually, the 3 targets move so close to each other which are being detected by both agents. When this happens, the local optimization of the tracking objective in Eqn. (\ref{eq:track_control}) directs each agent to track all 3 targets which causes the issue of tracking overlap. This is an unwanted behavior, which we wish to detect and resolve, in order to utilize the system resources for other tasks (e.g., searching).

In order to tackle this problem, instead of solving the joint tracking control problem which is a hard combinatorial problem that requires the enumeration of joint control actions among agents and the consideration of future multi-target states over a finite horizon, in this work we propose an alternative computationally cheaper way to tackle the tracking overlap problem. We allow any two agents to track the same targets but only for a short period of time. We consider that each agent can track multiple targets independent of other agents. When the trajectories of two or more tracking agents converge they exchange information to determine whether or not the exact same targets are being tracked. Once two agents have determined that they track exactly the same targets, one of them generates a search plan and switches to searching mode. The agent that switches to searching mode is picked at random, since each agent computes in this scenario approximately the same multi-target state as discussed next. The above procedure begins when two or more tracking agents have overlapping sensing ranges. \color{black}

Two agents $i$ and $j$ with states $s^i_{k-1}$ and $s^j_{k-1}$ respectively have overlapping sensing ranges when $\mathcal{S}_a(s^i_{k-1}) \cap \mathcal{S}_a(s^j_{k-1}) \neq \emptyset$ in which case the agents exchange their multi-target state estimates. Let the predicted multi-target states (regarding the true targets) of the agents $i$ and $j$ be $\hat{X}^{1,i}_{k|k-1}$ and $\hat{X}^{1,j}_{k|k-1}$, respectively. Also, let  $|\hat{X}^{1,i}_{k|k-1}| = m$ and $|\hat{X}^{1,j}_{k|k-1}| = n$ denote their cardinalities, i.e., the number of predicted targets in the set, with $n \ge m$ and $n, m \ne 0$. When $\mathcal{S}_a(s^i_{k-1}) \cap \mathcal{S}_a(s^j_{k-1}) \neq \emptyset$, the agents exchange their predicted multi-target states to compute the \textit{incremental tracking overlap score} as:
\begin{align} \label{eq:ospa}
&\Delta L^c_{k}(\hat{X}^{1,i}_{k|k-1},\hat{X}^{1,j}_{k|k-1}) = \notag \\
&\>\>\> \Bigg[~\frac{1}{n} \Bigg(\underset{\pi \in \Pi_n}{\text{min}} ~\underset{l=1}{\sum^m} ~ d_c(x^i_l,x^j_{\pi(l)})^2 + (n-m) \cdot c^2 \Bigg) ~ \Bigg]^{\frac{1}{2}}
\end{align}
where $x^i \in \hat{X}^{1,i}_{k|k-1}$, $x^j \in \hat{X}^{1,j}_{k|k-1}$ and $\Pi_n$ denotes the set of all permutations of size $m$ taken from the set $\{1,2,...,n\}$. The function $d_c(x,y) = \text{min}(c, \normvec{x-y}_2)$ where the parameter $c>0$ penalizes the cardinality mismatch between two sets. When $n<m$ Eq. (\ref{eq:ospa}) becomes $\Delta L^c_{k}(\hat{X}^{1,j}_{k|k-1},\hat{X}^{1,i}_{k|k-1})$. The above equation is called the optimal sub-pattern assignment (OSPA) \cite{Schuhmacher2008} of order 2. Then the \textit{cumulative tracking overlap score} for the time-window $[\kappa:K]$ is then defined as $Q_{\kappa:K}(s^i_{\kappa-1},s^j_{\kappa-1}) =
\sum_{k=\kappa}^{K} \mathcal{I}(\mathcal{S}_a(s^i_{k-1}),\mathcal{S}_a(s^j_{k-1}))  \cdot \Delta L^c_{k}(\hat{X}^{1,i}_{k|k-1},\hat{X}^{1,j}_{k|k-1})$
\noindent  where the function $\mathcal{I}(A,B)$ checks if the intersection of two regions $A$ and $B$ is non-empty and returns $1$, otherwise returns a large finite penalty. The cumulative tracking overlap score will generate a low score if two agents track the exact same targets over a certain period of time. In other words, when two agents have overlapping sensing ranges and they track the same number of targets with small positioning errors, the cumulative tracking overlap score is minimized. In order to determine if there is tracking overlap between two agents over a time-window the cumulative tracking overlap score is compared against a pre-determined threshold $Q^{Th}$. If $Q_{\kappa:K} \le Q^{Th}$ then the two agents track with high certainty the exact same targets. When this happens one of the two agents generates a search plan and switches to searching in the next time-step.

\section{Evaluation}
\label{sec:Evaluation}
\subsection{Experimental Setup}

In our experimental setup we assume that the targets maneuver in an area of 100m $\times$ 100m and that the single target state at time $k$ is described by $x_k = [\text{x},\dot{\text{x}},\text{y},\dot{\text{y}}]^\top$ i.e. position and velocity components in $xy$-direction. The target dynamics are piecewise linear according to the near constant velocity model with the process noise being Gaussian.  The single target transitional density is given by $\pi(x_k|x_{k-1}) = \mathcal{N}(x_k;Fx_{k-1},Q)$ where:
\begin{equation*}
	F \!=\! \begin{bmatrix} 
        1 & T & 0 & 0\\
        0 & 1 & 0 & 0\\
        0 & 0 & 1 & T\\
        0 & 0 & 0 & 1\\
       \end{bmatrix}, 
    Q = \sigma_w^2 \begin{bmatrix} 
        T^3/3 & T^2/2 & 0 & 0\\
        T^2/2 & T & 0 & 0\\
        0 & 0 & T^3/3 & T^2/2\\
        0 & 0 & T^2/2 & T\\
       \end{bmatrix}
\end{equation*}
\noindent with sampling interval $T=1$s and process noise intensity $\sigma_w^2 = 1 \text{ m}^2/\text{s}^3$. The target speed is initialized to $1$m/s in the $x$-direction and $1$m/s in the $y$-direction. The target survival probability from time $k-1$ to time $k$ is constant $p_{s,k}(x_{k-1})=0.99$ and does not depend on the target's state. Once an agent detects a target it receives range and bearing measurements thus the measurement model is given by $h_k(x_k,s_k) = \left[\norm{s_k -\text{H}x_k}_2,~ \text{atan2}\left(\text{y} - s_y, \text{x} - s_x \right) \right]^\top$ where $\text{H}$ is a matrix which extracts the target position from its state vector. The single target likelihood function is then given by $g(z_k|x_k,s_k) = \mathcal{N}(z_k;h_k(x_k,s_k),\Sigma^\top \Sigma)$ and $\Sigma$ is defined as $\Sigma = \text{diag}(\sigma_\zeta,\sigma_\phi)$. The standard deviations $(\sigma_\zeta,\sigma_\phi)$ are range dependent and given by 	$\sigma_\zeta = \zeta_0 + \beta_\zeta \norm{s_k-\text{H}x_k}_2^2$ and $\sigma_\phi = \phi_0 + \beta_\phi \norm{s_k-\text{H}x_k}_2 $ respectively with $\zeta_0 = 1$m, $\beta_\zeta = 5\times10^{-5}\text{m}^{-1}$, $\phi_0 = \pi/180$rad and $\beta_\phi=10^{-5}\text{rad}/\text{m}$. 
Moreover, the agent receives spurious measurements (i.e. clutter) with fixed Poisson rate $\lambda_c = 10$ uniformly distributed over the measurement space. Target births are distributed inside the agent sensing area with average rate of 3 births per time-step.
The agent's sensing model parameter $p_D^{\text{max}} = 0.99$ and the agent sensing area is $\mathcal{S}_{10}(s_k) = 10^2 ~\text{m}^2$.
The agent's dynamical model has radial displacement $\Delta_R=2$m, $N_R=2$ and $N_\theta = 8$ which gives a total of 17 control actions, including the initial position of the agent. Without loss of generality we assume in this evaluation that the function $J_k(x)$ is constant, state independent and equal to $J_k(x) = 0.999~ \forall x, k$.  
The agent communication range is $C_R=50$m, the Renyi divergence parameter is set to $\alpha = 0.5$ and the tracking overlap threshold is $Q^{Th} = 0.9$m during a time-window of length 3.

Finally, we should point out that in our implementation we have used the Sequential Monte Carlo (SMC) version of the PHD filter \cite{phd_1} for which the convergence properties have been established in \cite{phd_2}. Please note that under the linear, Gaussian assumptions on the target dynamics and measurement model the more efficient Gaussian-mixture \cite{gmphd} (GM-PHD) implementation of the PHD filter can be used which does not require the computationally expensive clustering step of the SMC-PHD filter, which is  used to partition the multi-target particle system into distinct target tracks. For mild non-linearities the GM-PHD filter can also be utilized by approximating the Gaussian-mixture using the extended and unscented Kalman filters. However, the severe limitation of the GM-PHD filter is that it requires a constant probability of detection (i.e., $p_D(x) = p_D$) and a constant target survival probability (i.e., $p_S(x) = p_S$). Lastly, more recently an alternative formulation of the SMC-PHD filter \cite{phd_3} has been proposed which avoids the particle system clustering step.
\color{black}

\subsection{Results}

\begin{figure}
	\centering
	\includegraphics[scale=0.4]{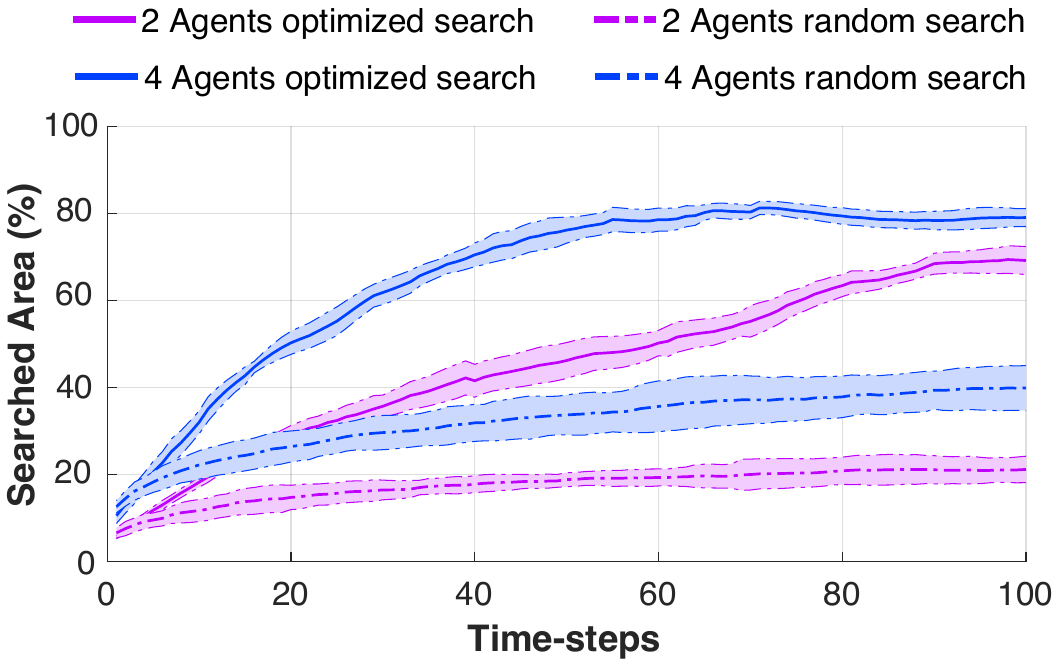}
	\caption{The figure shows the performance of the cooperative multi-agent path-planning and searching technique proposed in this work versus a random search scheme.}
	\label{fig:fig6}
	\vspace{-6mm}
\end{figure}

First we compare the performance of the proposed cooperative searching technique against a baseline random search scheme. To be more specific in this random search scheme a) no path-planning is performed by the agents i.e., the search-planning technique discussed in subsection \ref{ssec:cooperative_search} is not applied and b) there is no communication between the agents and thus no joint search-plans are produced i.e., Alg. \ref{alg:oppsearch} is not applied. Instead each agent, randomly selects and applies an admissible control action $u^k \in \mathbb{U}^{k}$ towards the unvisited regions. On the other hand the proposed technique performs a coordinated search-planning with $C_R=50$m.

We have conducted 50 Monte Carlo trials where we have randomly initialized the agents (uniformly distributed) inside the surveillance area and let the system run for 100 time-steps, measuring the percentage of searched area over time. This procedure is conducted for the proposed system and for the baseline random search scheme. Figure \ref{fig:fig6} shows the results of this experiment for the case of 2 and 4 agents. The figure shows that as the number of agents increases the searching performance increases as well which is reasonable. In addition the proposed optimized cooperative search approach significantly improves with the number of agents and outperforms the baseline method. This is due to the more efficient coordinated search planning and the communication between the agents. We should note here that the searched area in this experiment never reaches the 100\%. This is due to the decay of the searched density over time. Different decay rate and/or number of agents will result in different results. 

\begin{figure}
	\centering
	\includegraphics[width=\columnwidth]{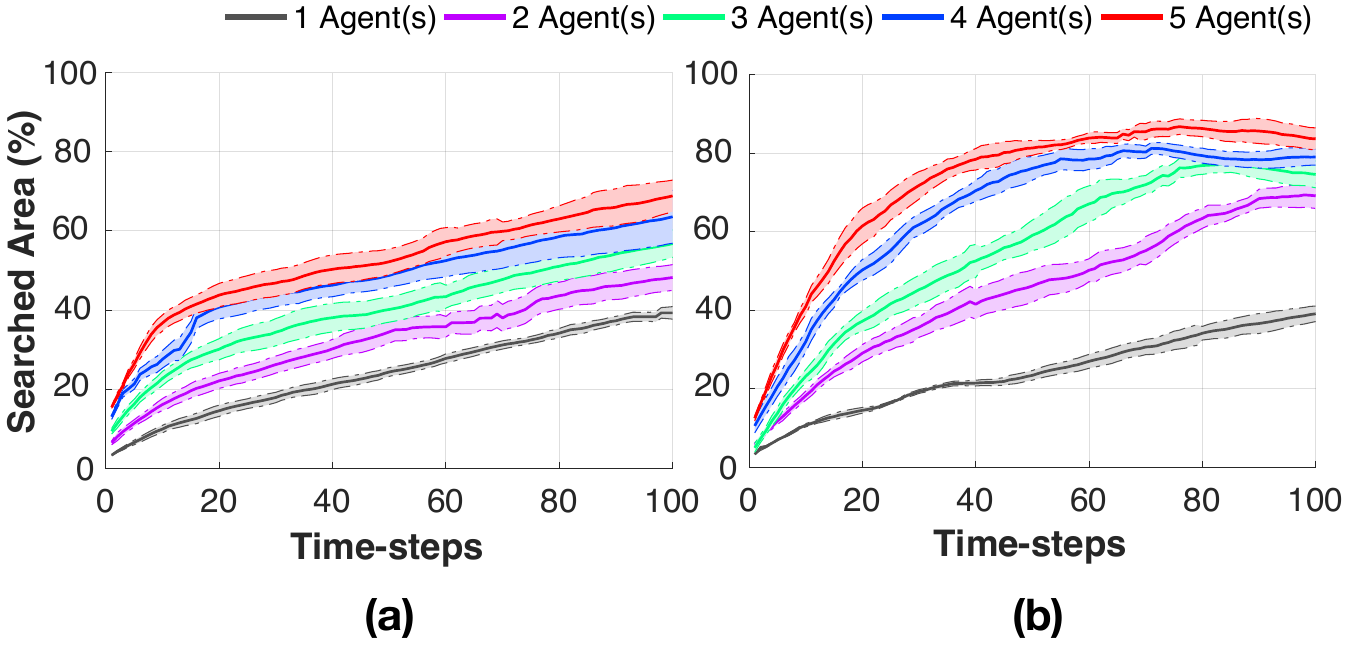}
	\caption{The figure illustrates the impact of communication range on the performance of the cooperative multi-agent searching. (a) Communication range $C_R=10$m, (b) Communication range $C_R=50$m.}
	\label{fig:fig5}
	\vspace{-6mm}
\end{figure}

The next experiment demonstrates the impact of the communication range $(C_R)$ on the performance of the proposed cooperative multi-agent searching approach.
Two different values of the communication range i.e. $C_R=10$m and $C_R$ = 50m are investigated. More specifically, Fig. \ref{fig:fig5} shows the percentage of searched area over time during 100 time-steps for $C_R=10$m (Fig. \ref{fig:fig5}a) and $C_R=50$m (Fig. \ref{fig:fig5}b). The figure shows the average value obtained from 50 MC trials with the agents randomly spawned inside the surveillance area. As we can observe from the figure, the increased communication range between the agents significantly improves the performance of the proposed approach. This is because the agents compute search paths cooperatively and thus the search task is solved jointly. 

In the next experiment, we study the joint search-and-track behavior of the proposed system and we use the optimal sub-pattern assignment (OSPA) error \cite{Schuhmacher2008} to quantify its performance. The OSPA metric is defined as the distance between two sets of points. It is used to jointly characterize the dissimilarity in the number of points and the values of the points in the respective sets. Since the output of the tracking approach utilized in this work is an estimated set of points in each time-step (i.e., the targets being tracked), OSPA metric will give us the deviation of this estimated set of points from the ground-truth set of points.  

\begin{figure}
	\centering
	\includegraphics[width=\columnwidth]{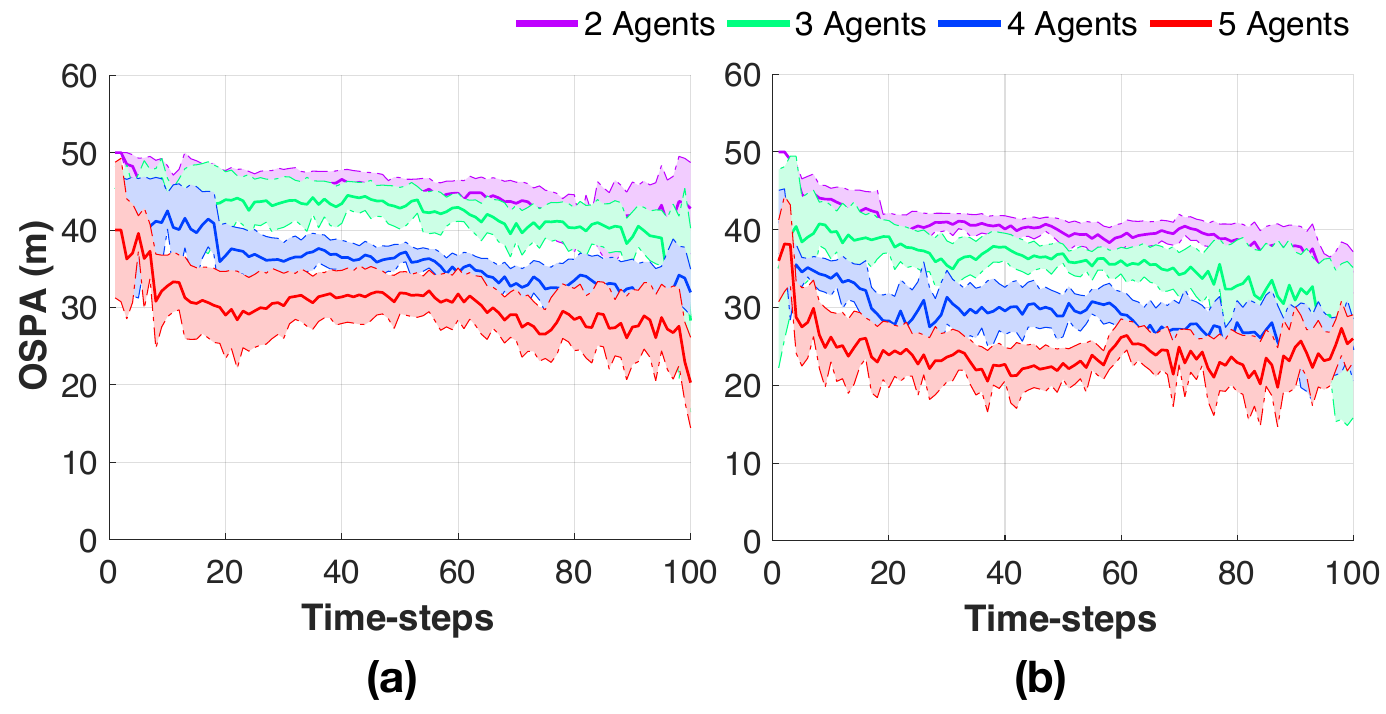}
	\caption{The figure illustrates the search-and-track performance of the proposed system by means of the OSPA error for two configurations of the communication range: (a) $C_R=10$m, (b) $C_R=50$m.}
	\label{fig:fig7}
	\vspace{-4mm}
\end{figure}

More specifically, for this experiment 10 targets are randomly spawned from the center of the surveillance area with random headings, following a nearly constant-velocity motion model. The average target life-time is 60 time-steps. At time-step $k=1$ we uniformly distribute a fixed number of agents inside the surveillance area and we let the system run for 100 time-steps. Figure \ref{fig:fig7} shows the average OSPA error (OSPA order=2, $c=50$m) over 50 MC trials for 2, 3, 4 and 5 agents operating with communication ranges of $C_R=10$ in Fig. \ref{fig:fig7}a, and $C_R=50$ in Fig. \ref{fig:fig7}b. We first observe that the tracking accuracy of the proposed multi-agent system increases with the number of agents. The figure shows that the OSPA error starts high but subsequently decreases. This is because initially the agents start in search mode however, when at some point a target is detected the respective agent switches to tracking mode which results in a decrease in the OSPA error. 
Finally, we observe that the increased communication range results in improved multi-agent cooperative searching which in turn results in better target discovery and thus improved tracking accuracy, as shown in Fig. \ref{fig:fig7}b.

\begin{figure}
	\centering
	\includegraphics[width=\columnwidth]{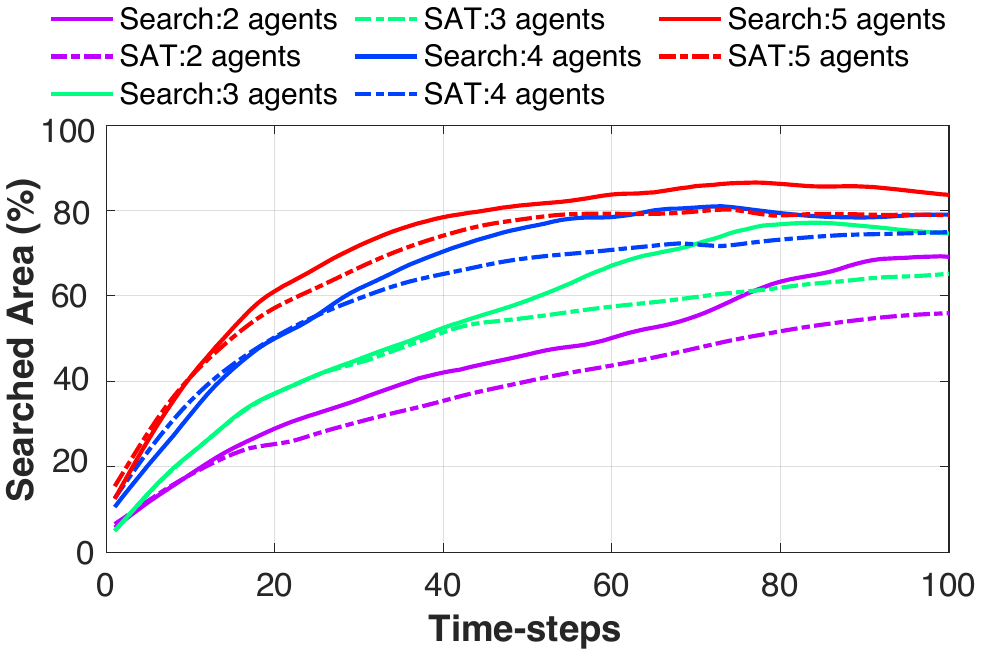}
	\caption{The figure shows the percentage of the searched area for the tasks of cooperative searching versus cooperative search-and-track for the case of 2, 3, 4 and 5 agents.}
	\label{fig:fig8}
	\vspace{-6mm}
\end{figure}

For the next experiment, we followed a similar setup with the one described in the previous paragraph where in the first case (termed Search) a number of agents (i.e., 2, 3, 4 and 5) with $C_R=50$, are uniformly distributed inside the surveillance area. In the second case (termed SAT), in addition to the agents, we uniformly distribute 10 targets inside the surveillance area, with random headings and average life-time of 60 time-steps. We let the system run for 100 time-steps and we measure the percentage of searched area over time. Figure \ref{fig:fig8} shows the average percentage of the searched area over 60 MC trials. As we can observe, the searching performance drops during the searching-and-tracking (SAT) task compared to the Search task. When an agent detects the presence of targets, automatically switches to tracking mode which produces sub-optimal search results. One possibility to mitigate this, which is left for future work, would be with a target hand-over strategy i.e., the agents would hand over targets to their peers who have search plans that align better with respect to the target's trajectory. 


\begin{figure}
	\centering
	\includegraphics[width=\columnwidth]{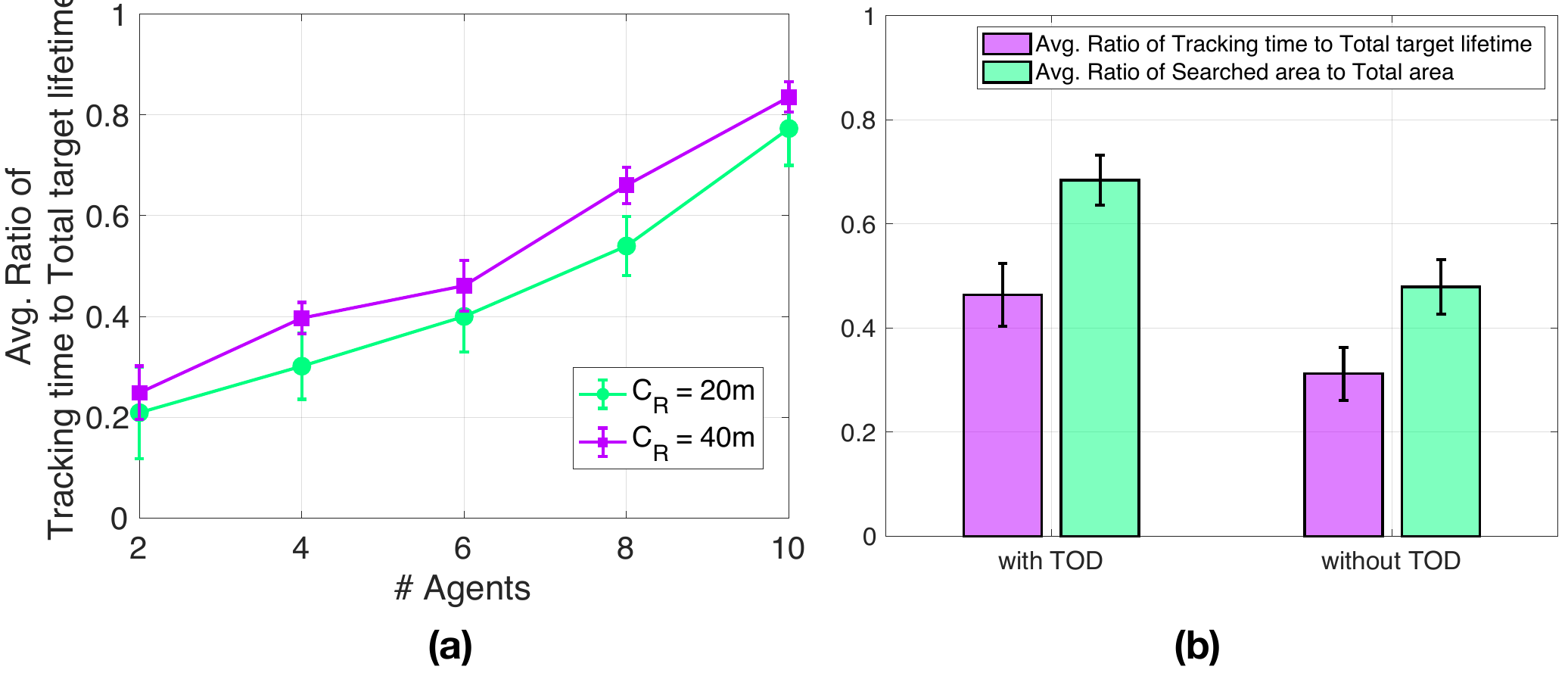}
	\caption{The figure shows a) the average ratio of target tracking time to target lifetime as a function of the number of agents and communication range b) the performance gain from the use of tracking overlap detection and resolution.}
	\label{fig:fig10}
	\vspace{-6mm}
\end{figure}

Our two final experiments aim to investigate a) the expected time for which a target can be tracked successfully in a multi-agent, multi-target setting and b) the performance improvement gained with the use of the tracking overlap detection and resolution approach discussed in Sec. \ref{ssec:multi_agent_tracking}(b).
In order to investigate the expected time for which a target can be tracked successfully we have used the following procedure: We first randomly generate 20 target trajectories inside the surveillance region with average lifetime of 30 time-steps. The target birth/death times vary as the targets enter and exit the surveillance region and their birth locations are uniformly distributed inside the whole area of $100\text{m} \times 100\text{m}$. The targets move with an average speed of $1$m/s in the $x$-direction and $1$m/s in the $y$-direction. The agents are uniformly distributed inside the surveillance area. We let the system run for 100 time-steps and we measure the time for which the agents successfully track the targets. This time is then normalized with the actual lifetime of the targets (i.e., the ground-truth length of time for which a target evolves inside the surveillance area). We conduct 50 Monte Carlo trials with 2, 4, 6, 8 and 10 agents and with communication ranges of 20m and 40m. The results of this experiment is shown in Fig. \ref{fig:fig10}a. As we can observe the expected tracking time increases with the number of agents and with the communication range. The average ratio of tracking time to the total target lifetime starts from around 0.25 for 2 agents and reaches 0.83 for 10 agents operating with $C_R=40$m. A similar trend is also true for the communication range of 20m. In this setup, the increase in the communication range increases the search efficiency through cooperation which as a result improves the expected tracking time. 
Finally, in order to investigate the performance gain from the proposed tracking overlap detection and resolution approach we have conducted a similar experimental setup (i.e., same parameters as before unless otherwise noted) in which we randomly spawn 15 targets and 5 agents and we fix the communication range to 20m. Figure \ref{fig:fig10}b shows a) the average ratio of tracking time to total target lifetime and b) the average ratio of searched area to total area over 50 Monte Carlo trials with and without tracking overlap detection (TOD). As we can observe the proposed tracking overlap detection and resolution approach increases both the multi-agent tracking accuracy and the searching performance of the system. This is because TOD allows the overlapping agents to disengage from tracking and switch to searching. As a result the area is searched more efficiently which in turn improves the target discovery and ultimately improves the tracking performance.

\section{Conclusion} \label{sec:Conclusion}
In this work a novel decentralized cooperative multi-agent searching-and-tracking framework has been proposed. The proposed approach recursively computes and propagates in time the \textit{searching-and-tracking} (SAT) density which is used by the agents to devise efficient cooperative searching and tracking strategies. The proposed framework is flexible and accounts for many of the challenges present in search and rescue missions including the unknown and time varying number of targets, the noisy sensor measurements, the uncertain target dynamics and the limited sensing range of the agents. In the future we plan to investigate in more detail the communication aspects (e.g., the communication overhead vs performance) of the proposed approach. Although, the event-based strategy suggested in this work can reduce the communication overhead of the searching-and-tracking task, it sacrifices the overall performance of the team, since no global coordination is guaranteed. Future work aims to investigate how communication-efficient cooperation \cite{Leung2010} can be enabled for the problem tackled in this work and how efficient distributed transmission protocols \cite{Liu2017d} can be utilized in order to further reduce the communication burden between the agents. \color{black}

\section*{Acknowledgments}
This work is supported by the European Union Civil Protection under grant agreement  No 783299  (SWIFTERS), by the European Union’s Horizon 2020 research and innovation programme under grant agreement No 739551 (KIOS CoE) and from the Republic of Cyprus through the Directorate General for European Programmes, Coordination and Development.

\balance
\bibliographystyle{IEEEtran}
\bibliography{IEEEabrv,main} 

\end{document}